\documentclass[12pt]{article}%
\usepackage{makeidx}
\usepackage{amsmath}
\usepackage{amssymb}
\usepackage{geometry}
\usepackage{mathptmx}
\usepackage{amsfonts}
\usepackage{graphicx}%
\providecommand{\U}[1]{\protect\rule{.1in}{.1in}}

\begin{document}

\begin{center}
\textbf{Superconductor Meissner effects for gravito-electromagnetic fields in
harmonic coordinates due to non-relativistic gravitational sources}%
\vspace{0.15in}

{\large Nader A. Inan}\footnote{email: ninan@ucmerced.edu
\par
{}}

Clovis Community College,

10309 N. Willow, Fresno, CA 93730, USA\vspace{0.3in}

\textbf{Abstract}\vspace{0.15in}
\end{center}

It is well known that a covariant Lagrangian for relativistic charged
particles can lead to a vanishing Hamiltonian. Alternatively, it is shown that
using a \textquotedblleft space+time\textquotedblright\ Lagrangian leads to a
new canonical momentum and minimal coupling rule that describes the coupling
of both electromagnetic and gravitational fields to a relativistic charged
particle. Discrepancies between Hamiltonians obtained by various authors are
resolved. The canonical momentum leads to a new form of the London equations
and London gauge. Using the linearized Einstein field equation in harmonic
coordinates, and a non-relativistic ideal fluid, leads to
gravito-electromagnetic field equations. These are used to obtain new
penetration depths for both the magnetic and gravito-magnetic fields. A key
result is that the gravito-magnetic field is expelled from a superconductor
only when a magnetic field is also present. The flux quantum in the body of a
superconductor, and the quantized supercurrent in a superconducting ring are
derived. Lastly, the case of a superconducting ring in the presence of a
charged rotating mass cylinder is used as an example of applying the formalism
developed.\vspace{0.3in}

\noindent\underline{\textbf{Introduction}}\bigskip

In 1949, the Gravity Research Foundation (GRF)\ was founded by Roger Babson
with the goal of finding practical applications of gravity such as partial
insulators, reflectors, or absorbers of gravity. \textquotedblleft His views
were reflected by the wording in the announcement of the first essay
competition that said the awards were to be given for suggestions for
anti-gravity devices, for partial insulators, reflectors, or absorbers of
gravity, or for some substance that can be rearranged by gravity to throw off
heat - although not specifically mentioned in the announcement, he was
thinking of absorbing or reflecting gravity waves.\textquotedblright%
\cite{GRF}\vspace{0.3in}

Yet by 1953, the winning GRF essay by Bryce DeWitt dispensed with such
endeavors as unrealizable. DeWitt expressed doubt that a material which
absorbs or reflects gravitational fields exists.\cite{DeWitt essay} He states,
\textquotedblleft\ldots first fix our sights on those grossly practical
things, such as `gravity reflectors' or `insulators', or magic `alloys' which
can change `gravity' into heat, which one might hope to find as the usual
by-products of new discoveries in the theory of gravitation\ldots\ Of primary
importance is the extreme weakness of gravitation coupling between material
bodies \ldots\ The weakness of this coupling has the consequence that schemes
for achieving gravitational insulation, via methods involving fanciful devices
such as oscillation or conduction, would require masses of planetary
magnitude.\textquotedblright\pagebreak

However, in 1966 DeWitt's viewpoint changed when he published an influential
paper predicting the Lense-Thirring field (also known as frame-dragging or
gravito-magnetic field) is expelled from superconductors in a Meissner-like
effect. \cite{DeWitt} It is evident DeWitt considered quantum mechanical
systems (such as superconductors) may possess characteristics that were not
previously considered when the previous statement was made barring the
possibility of gravitational reflectors or insulators.\vspace{0.3in}

In \cite{DeWitt}, DeWitt begins with the Lagrangian for a relativistic charged
particle in an electromagnetic field in curved space-time and develops the
associated Hamiltonian. He identifies a minimal coupling rule involving a
gravitational vector potential and concludes this result implies the
associated gravitational field must be expelled from a superconductor, just as
the magnetic field is expelled from a superconductor in the Meissner
effect.\vspace{0.3in}

DeWitt's novelty and intuition is commendable, however, there are some
technical shortcomings in his treatment. Also, his interpretation of the flux
quantization condition, and his order of magnitude calculation for an induced
electric current, are questioned here. In particular, the following items will
be demonstrated.\bigskip

\begin{enumerate}
\item The Hamiltonian formulated by DeWitt (shown below) contains errors.%
\begin{align}
H_{DeWitt}  &  =c\left(  g^{jk}g_{0j}g_{0k}-g_{00}\right)  ^{1/2}\left[
m^{2}c^{2}+g^{jk}\left(  P_{j}-eA_{j}\right)  \left(  P_{k}-eA_{k}\right)
\right]  ^{1/2}\nonumber\\
&  -cg^{jk}g_{0k}\left(  P_{j}-eA_{j}\right)  -ceA_{0} \label{H_DeWitt}%
\end{align}

\item The weak field, low velocity limit of the Hamiltonian is missing a
critical term. DeWitt reduced the Hamiltonian to%
\begin{equation}
H_{DeWitt}=\frac{1}{2m}\left(  \vec{P}-e\vec{A}-m\vec{h}_{0}\right)  ^{2}+V
\label{H_DeWitt '}%
\end{equation}
where%
\begin{equation}
V=-eA_{0}-\tfrac{1}{2}mh_{00}\qquad\text{and}\qquad\vec{h}_{0}=c\left(
h_{01,}h_{02,}h_{03}\right)  \label{B_G field}%
\end{equation}
Here $e$ and $m$ are the charge and mass of an electron, respectively. The
Hamiltonian includes a term involving $m\vec{h}^{2}$ which is inconsistent
with a linear approximation in the metric perturbation. The Hamiltonian is
also missing several terms that are of comparable magnitude as the ones retained.

\item The minimal coupling rule, $\vec{P}\rightarrow\vec{P}-e\vec{A}-m\vec
{h}_{0}$ is missing several terms of comparable magnitude. The missing terms
impact the associated London equations, London gauge, and penetration depths
for the magnetic field and the gravito-magnetic field. A careful treatment
will show that the gravito-magnetic field is only expelled while a magnetic
field is also present. In the absence of a magnetic field, the superconductor
exhibits a paramagnetic effect rather than a diamagnetic (Meissner) effect for
the gravito-magnetic field.

\item The gravito-magnetic field used in DeWitt's formulation is
coordinate-dependent. With an appropriate coordinate transformation, the field
can be made to vanish. Therefore, the corresponding gravitational
Meissner-like effect is also coordinate-dependent and can be made to vanish as well.

\item DeWitt argues that the flux of the vector $\vec{G}=e\nabla\times\vec
{A}+m\nabla\times\vec{h}_{0}$ must be quantized in units of $h/2$. In
actuality, there is an additional term involving the flux of $\vec{E}%
_{G}\times\vec{A}$ that should be included.

\item Lastly, DeWitt claims that a magnetic field must arise if a
superconducting ring is concentric with a rotating massive cylinder which is
producing a gravito-magnetic flux in the ring. He predicts an electric current
will be induced with an order-of-magnitude given by%
\begin{equation}
I\sim\frac{GmMV}{ed} \label{I (magnitude)}%
\end{equation}
where $V$ is the rim velocity, $d$ is the diameter, and $M$ is the mass of the
rotating cylinder. However, it will be shown that the electric current DeWitt
predicted is actually limited to $I<h/\left(  2eL\right)  $, where $L$ is the
self-inductance. Furthermore, it will be argued that if there is no external
magnetic field present, then the flux of the gravito-magnetic field will be
quantized, but this will not induce the flux of a magnetic field.\vspace
{0.3in}
\end{enumerate}

In Section I, we begin with a covariant Lagrangian for charged relativistic
test particles in an electromagnetic field in curved space-time. The
Euler-Lagrange equation of motion leads to the geodesic equation of motion
modified by the Lorentz four-force in curved space-time. Although the equation
of motion correctly describes the dynamics of the particle, the associated
Hamiltonian is identically zero and therefore cannot be used to describe a
quantum mechanical system such as a superconductor.\vspace{0.3in}

Alternatively, a \textquotedblleft space+time\textquotedblright\ Lagrangian is
used to obtain a canonical three-momentum and Hamiltonian valid to all orders
in the metric. The result is compared to DeWitt's in $\left(  \ref{H_DeWitt}%
\right)  $. Some relevant metric relationships will be used to check the
Hamiltonian with that of other authors for confirmation of its validity. The
Hamiltonian is then expanded to first order in the metric to show the lowest
order coupling of the momentum, electromagnetic fields, and gravitational
fields. Again, the result is compared to DeWitt's result in $\left(
\ref{H_DeWitt '}\right)  $. The Hamiltonian is further simplified by
introducing the trace-reversed metric perturbation and assuming
non-relativistic gravitational sources.\vspace{0.3in}

In Section II, gravito-electric and gravito-magnetic fields are defined in
terms of the metric perturbation. Using the stress tensor of a
non-relativistic ideal fluid, and the linearized Einstein field equation in
harmonic coordinates, leads to gravito-electromagnetic field equations. In
addition, the canonical three-momentum is used to develop constitutive
equations for the supercurrent. These lead to a new set of London equations
describing the interaction of electromagnetic and gravito-electromagnetic
fields with a superconductor. A modification to the London gauge condition is
also identified.\vspace{0.3in}

Next, the constitutive equations are used in the field equations to identify a
penetration depth associated with the magnetic field and the gravito-magnetic
field. It is found that the usual London penetration depth for the magnetic
field is modified by the presence of a gravito-magnetic field, however, the
modification is miniscule. It is also found that in the absence of a magnetic
field, the superconductor demonstrates a paramagnetic effect rather than a
diamagnetic (Meissner)\ effect for the gravito-magnetic field. In other words,
the gravito-magnetic field is not expelled. However, when the magnetic field
and the gravito-magnetic field are \textit{both} present, then it is found
that the gravito-magnetic field is expelled with a penetration depth on the
same order as the London penetration depth. However, it is demonstrated that
the gravito-magnetic field is a coordinate-dependent quantity and therefore
effects associated with it can be made to vanish with an appropriate
coordinate transformation.\vspace{0.3in}

In Section III, the new minimal coupling rule obtained in Section I\ will be
used to write the Ginzburg-Landau supercurrent with coupling to
electromagnetic and gravitational fields. Using the fact that all the fields
vanish within the body of the superconductor, and the complex order parameter
must be single-valued around a closed path, leads to a quantization condition
for the flux of the magnetic and gravitational fields. The quantization
condition is compared to DeWitt's result which states the flux of the vector
$\vec{G}=e\nabla\times\vec{A}+m\nabla\times\vec{h}_{0}$ must be quantized in
units of $h/2$.\vspace{0.3in}

Lastly, in Section IV, the canonical momentum is used to develop an expression
for the Ginzburg-Landau phase around the perimeter of a superconducting ring.
Once again, using the fact that the wave function is single-valued leads to an
extension of the Byers-Yang theorem to the case of electromagnetic and
gravitational fields. The result is a quantization condition involving the
flux of the magnetic and gravitational fields, and the supercurrent around the
perimeter of a superconducting ring. A charged, rotating mass cylinder is
introduced as a source for electromagnetic and gravitational fields. The
effect of placing the rotating cylinder concentric with the superconducting
ring is analyzed. It is argued that the electric current predicted by DeWitt
in $\left(  \ref{I (magnitude)}\right)  $ does not occur. Rather, any
pre-existing supercurrent in the ring is quantized along with the flux of
electromagnetic and gravitational fields through the ring.\vspace
{0.3in}\vspace{0.3in}

\noindent\underline{\textbf{I. A Hamiltonian for Cooper pairs coupled to
electromagnetism in curved space-time}}\bigskip

Using an action of the form $S=\int_{\tau_{1}}^{\tau_{2}}Ld\tau$, where $\tau$
is proper time, leads to a Lagrangian for charged relativistic test particles
in an electromagnetic field in curved space-time that can be written
as\footnote{The signature of the Minkowski metric used here is
diag$(-1,+1,+1,+1)$. Greek space-time indices $\mu,\nu,...$ run from 0 to 3.
Latin spatial indices $i,j,...$ run from 1 to 3.
\par
{}}%
\begin{equation}
L=-mc\sqrt{-g_{\mu\nu}u^{\mu}u^{\nu}}-qg_{\mu\nu}u^{\mu}A^{\nu}
\label{4-velocity Lagrangian (4-vectors)_}%
\end{equation}
where $u^{\mu}=dx^{\mu}/d\tau$ is the four-velocity, $A^{\mu}=\left(
\varphi/c,A^{i}\right)  $ is the electromagnetic four-potential, and $m$ and
$q$ are the rest mass and charge of the test particles, respectively. It is
well known that using $\left(  \ref{4-velocity Lagrangian (4-vectors)_}%
\right)  $ in the Euler-Lagrange equation of motion%
\begin{equation}
\dfrac{\partial L}{\partial x^{\mu}}-\frac{d}{d\tau}\dfrac{\partial L}{\left(
\partial x^{\mu}/d\tau\right)  }=0
\end{equation}
leads to the geodesic equation of motion modified by the Lorentz four-force in
curved space-time%
\begin{equation}
\dfrac{dp^{\mu}}{d\tau}+m\Gamma_{\sigma\rho}^{\mu}u^{\sigma}u^{\rho}%
=qg_{\nu\alpha}u^{\alpha}F^{\mu\nu}%
\end{equation}
where $\Gamma_{\sigma\rho}^{\mu}$ are the metric connections (Christoffel
symbols).\footnote{See Box 13.3 in MTW\ \cite{MTW} or 3.3 of Wald \cite{Wald}.
\par
{}} This demonstrates that the Lagrangian in $\left(
\ref{4-velocity Lagrangian (4-vectors)_}\right)  $ correctly characterizes the
dynamics in a covariant form. However, evaluating a canonical four-momentum,
$P_{\mu}=\partial L/\partial u^{\mu}$, leads to $mu^{\mu}=P^{\mu}+qA^{\mu}$,
where $q=-e$ \ for electrons. This implies a minimal coupling rule given by
$P^{\mu}\rightarrow P^{\mu}-eA^{\mu}$. It is well known that applying this
minimal coupling rule, and utilizing a covariant Legendre transformation,
$H=P_{\mu}u^{\mu}-L$, leads to a Hamiltonian that is identically
zero.\footnote{This is mentioned in Chapter 12 of Jackson \cite{Jackson} for
flat space-time, and discussed in more detail in \cite{Bertschinger (GL)},
\cite{Cisneros-Parra}, and \cite{Cognola} for curved space-time. It is also
shown in \cite{Bertschinger} that a covariant Lagrangian of the form
$L_{2}=\dfrac{1}{2m}g_{\mu\nu}p^{\mu}p^{\nu}+eA_{\mu}u^{\mu}$ will lead to a
non-vanishing Hamiltonian, $H=\dfrac{1}{2m}\left(  P^{\mu}-eA^{\mu}\right)
^{2}$. Using a \textquotedblleft space+time\textquotedblright\ approach leads
to the Hamiltonian shown in $\left(  \ref{CVZ}\right)  $.
\par
{}} Therefore, it is evident that a different approach must be taken for
obtaining a Hamiltonian as will be shown next.\vspace{0.3in}

Note that the four-velocity can be written as $u^{\mu}=\gamma v^{\mu}$, where
the Lorentz factor\ is $\gamma\equiv dt/d\tau$, the coordinate velocity is
$v^{\mu}=dx^{\mu}/dt=\left(  c,v^{i}\right)  $, and $t$ is the coordinate
time. The action can also be written as $S=\int_{\tau_{1}}^{\tau_{2}}%
Ld\tau=\int_{t_{1}}^{t_{2}}L\gamma^{-1}dt$ which is now reparametrized in
terms of coordinate time rather than proper time. Therefore, the
\textquotedblleft space+time\textquotedblright\ Lagrangian
becomes\footnote{This is essentially the Lagrangian used by DeWitt
\cite{DeWitt} except he uses the notation $v^{\mu}=\dot{x}^{\mu}$ and sets
$c=1$. Most authors such as DeWitt \cite{DeWitt}, Cognola, et al.
\cite{Cognola} and Bertschinger \cite{Bertschinger} leave the electromagnetic
field in the Langrangian $\left(  \ref{4-velocity Lagrangian (3-vectors)_DW_}%
\right)  $ as $A_{\mu}x^{\mu}$ instead of $g_{\mu\nu}A^{\mu}x^{\mu}$. However,
this neglects the coupling of gravity to the electromagnetic field. As a
result, the term involving $g_{0i}A^{0}$ would not appear in the canonical
momentum $\left(  \ref{canonical momentum}\right)  $ or the Hamiltonian
$\left(  \ref{H_general}\right)  $.
\par
{}}%
\begin{equation}
L=-mc\sqrt{-g_{\mu\nu}v^{\mu}v^{\nu}}-qg_{\mu\nu}v^{\mu}A^{\nu}
\label{4-velocity Lagrangian (3-vectors)_DW_}%
\end{equation}
Using $g_{\mu\nu}u^{\mu}u^{\nu}=-c^{2}$, the Lorentz factor\ in curved
space-time can be evaluated as%
\begin{equation}
\gamma=\sqrt{-g_{\mu\nu}v^{\mu}v^{\nu}}=\left(  -g_{00}-2g_{0j}v^{j}%
/c-g_{ij}v^{i}v^{j}/c^{2}\right)  ^{-1/2} \label{gamma}%
\end{equation}
Then the canonical three-momentum, $P_{i}=\partial L/\partial v^{i}$, can be
found from $\left(  \ref{4-velocity Lagrangian (3-vectors)_DW_}\right)  $ to
be\footnote{In \cite{Hirakawa}, a canonical momentum is also proposed in the
form of $P_{i}=mg_{ik}u^{k}-eA_{i}$. However, since it was not formally
derived from a Lagrangian, it is missing all the other terms in $\left(
\ref{canonical momentum}\right)  $.}%
\begin{equation}
P_{i}=\gamma m\left(  cg_{0i}+g_{ij}v^{j}\right)  -q\left(  g_{0i}A^{0}%
+g_{ij}A^{j}\right)  \label{canonical momentum}%
\end{equation}
In the absence of electromagnetic and gravitational fields, this reduces to
$P_{i}=\gamma mv_{i}$. Using $g_{\mu\nu}=\eta_{\mu\nu}+h_{\mu\nu}$, where
$\eta_{\mu\nu}$ is the Minkowski metric of flat space-time, and $h_{\mu\nu}$
is a perturbation, leads to%
\begin{equation}
\gamma mv_{i}=P_{i}-\gamma m\left(  ch_{0i}+h_{ij}v^{j}\right)  +q\left(
A_{i}+h_{0i}A^{0}+h_{ij}A^{j}\right)  \label{canonical momentum '}%
\end{equation}
This implies that in the presence of electromagnetic and gravitational fields,
there is a modified minimal coupling rule given by%
\begin{equation}
P_{i}\quad\rightarrow\quad P_{i}-\gamma m\left(  ch_{0i}+h_{ij}v^{j}\right)
+q\left(  A_{i}+h_{0i}A^{0}+h_{ij}A^{j}\right)  \label{minimal coupling}%
\end{equation}
For an electron $\left(  q=-e\right)  $ in flat space-time $\left(  h_{\mu\nu
}=0\right)  $, the minimal coupling rule reduces to the usual $P_{i}%
\rightarrow P_{i}-eA_{i}$.\vspace{0.3in}

The first term in $\left(  \ref{canonical momentum}\right)  $ can be defined
as a relativistic kinetic momentum in curved space-time%
\begin{equation}
\pi_{i}\equiv\gamma m\left(  cg_{0i}+g_{ij}v^{j}\right)  \label{pie_i}%
\end{equation}
Notice that in flat space-time, this becomes the usual relativistic kinetic
momentum, $\pi_{i}=\gamma_{u}mv_{i}$ where $\gamma_{u}=\left(  1-v^{2}%
/c^{2}\right)  ^{-1/2}$. Using a Legendre transformation, $H=P_{k}v^{k}-L$,
requires solving $\left(  \ref{pie_i}\right)  $ for $v^{i}$ which necessitates
constructing the inverse of $g_{ik}$. This is shown in $\left(  \ref{g^tilda}%
\right)  $ of Appendix A to be $\tilde{g}^{ik}\equiv g^{ik}-g^{0i}%
g^{0k}/g^{00}$ so that $\tilde{g}^{jk}g_{ik}=\delta_{\hspace{0.05in}i}^{j}$.
Then the velocity and the Lorentz factor can be expressed, respectively, as%
\begin{equation}
v^{j}=\dfrac{\tilde{g}^{jk}\pi_{k}}{\gamma m}-c\tilde{g}^{jk}g_{0k}%
\qquad\text{and}\qquad\gamma=\dfrac{1}{mc}\sqrt{\dfrac{m^{2}c^{2}+\tilde
{g}^{ik}\pi_{j}\pi_{k}}{\tilde{g}^{jk}g_{0j}g_{0k}-g_{00}}}%
\end{equation}
Using $\left(  \ref{canonical momentum}\right)  $ and $\left(  \ref{pie_i}%
\right)  $, as well as $q=-e$, makes Hamiltonian become%
\begin{align}
H  &  =c\left(  \tilde{g}^{jk}g_{0j}g_{0k}-g_{00}\right)  ^{1/2}\left[
m^{2}c^{2}+\tilde{g}^{jk}\left(  P_{j}-eg_{0j}A^{0}-eg_{ij}A^{i}\right)
\left(  P_{k}-eg_{0k}A^{0}-eg_{ik}A^{i}\right)  \right]  ^{1/2}\qquad
\nonumber\\
&  \qquad-c\tilde{g}^{jk}g_{0k}\left(  P_{j}-eg_{0j}A^{0}-eg_{ij}A^{i}\right)
-ec\left(  g_{00}A^{0}+g_{i0}A^{i}\right)  \label{H_general}%
\end{align}
Comparing this Hamiltonian to DeWitt's result in $\left(  \ref{H_DeWitt}%
\right)  $, it is evident DeWitt uses $g^{jk}$ rather than $\tilde{g}^{jk}$.
However, it is shown in $\left(  \ref{g^tilda_jk to first order}\right)  $
that $\tilde{g}^{jk}\approx g^{jk}$ is only true to first order in the metric
perturbation. Also note that using the metric relations developed in $\left(
\ref{g^ik*g_0k}\right)  $ and $\left(  \ref{g^tilda^jk*g_oj^g_ok}\right)  $
leads to $\left(  \ref{H_general}\right)  $ taking a form that matches
\cite{Cognola} and \cite{Bertschinger}.%
\begin{equation}
H=c\left[  \dfrac{m^{2}c^{2}+\tilde{g}^{jk}\pi_{j}\pi_{k}}{-g^{00}}\right]
^{1/2}+c\dfrac{g^{0j}}{g^{00}}\pi_{j}-ec\left(  g_{00}A^{0}+g_{i0}%
A^{i}\right)  \label{CVZ}%
\end{equation}
Staying to first order in the metric perturbation and assuming
non-relativistic velocities makes the Hamiltonian become\footnote{Note that
when working to first order in the metric, spatial indices can be freely
raised and lowered.}%
\begin{equation}
H=mc^{2}+\frac{\pi^{2}}{2m}-\dfrac{1}{2}h_{00}mc^{2}-h_{00}\frac{\pi^{2}}%
{4m}-ch_{0i}\pi^{i}+\frac{h_{ij}\pi^{i}\pi^{j}}{2m}+ecA^{0}-ech_{00}%
A^{0}-ech_{0i}A^{i} \label{H in terms of pi}%
\end{equation}
Using $\left(  \ref{canonical momentum}\right)  $ and $\left(  \ref{pie_i}%
\right)  $ gives%
\begin{equation}
\pi_{i}=P_{i}-eh_{0i}A^{0}-eA_{i}-eh_{ik}A^{i} \label{pie}%
\end{equation}
\pagebreak

\noindent Inserting $\left(  \ref{pie}\right)  $ into $\left(
\ref{H in terms of pi}\right)  $ makes the term involving $h_{0i}A^{i}$
cancel. The Hamiltonian is then expressed in terms of the canonical momentum
(to first order in the metric perturbation). Using $P^{2}\equiv P^{i}P^{i}$,
we have%
\begin{align}
H  &  =mc^{2}+\dfrac{1}{2m}\left(  P^{i}-eA^{i}\right)  ^{2}+e\varphi
\nonumber\\
& \nonumber\\
&  -\tfrac{1}{2}h_{00}mc^{2}-h_{00}\dfrac{P^{2}}{4m}-ch_{0i}P^{i}+\dfrac
{1}{2m}h_{ij}P^{i}P^{j}\nonumber\\
& \nonumber\\
&  -eh_{00}\varphi+\dfrac{e^{2}}{2m}\left(  3h_{ij}A^{i}A^{j}-\dfrac{1}%
{2}h_{00}A^{2}\right)  +\dfrac{e}{2m}\left(  h_{00}A^{i}+\dfrac{1}{c}\varphi
h_{0i}-4h_{ij}A^{j}\right)  P^{i} \label{H}%
\end{align}
The Hamiltonian consists of the following terms.

\begin{itemize}
\item The first line contains the standard terms for a charged particle
coupled to electromagnetic fields. (No coupling to gravity.)

\item The second line describes the coupling of the gravitational potentials
to the \textit{mass} and canonical momentum of the test particles. (No
coupling to electromagnetism.)

\item The third line describes the coupling of the electromagnetic and
gravitational potentials together to the \textit{charge} and canonical
momentum of the test particles.\bigskip
\end{itemize}

Note that the Hamiltonian contains the scalar, vector, and tensor parts of the
metric perturbation which are, respectively, $h_{00}$, $h_{0i}$, and $h_{ij}$.
In particular, the tensor part is still first order in the metric perturbation
but is missing in DeWitt's Hamiltonian $\left(  \ref{H_DeWitt '}\right)  $. To
eliminate the tensor part, we can use the trace-reversed metric perturbation
defined as $\bar{h}_{\mu\nu}\equiv h_{\mu\nu}-\tfrac{1}{2}\eta_{\mu\nu}h$,
where $h=\eta^{\mu\nu}h_{\mu\nu}$. It will be shown in the following section
that using the harmonic coordinate condition, $\partial^{\nu}\bar{h}_{\mu\nu
}=0$, leads to $\bar{h}_{ij}=0$ for non-relativistic gravitational sources.
Then using $h_{\mu\nu}=\bar{h}_{\mu\nu}-\tfrac{1}{2}\eta_{\mu\nu}\bar{h}$,
leads to%
\begin{equation}
h_{00}=\tfrac{1}{2}\bar{h}_{00},\qquad h_{0i}=\bar{h}_{0i},\qquad
h_{ij}=\tfrac{1}{2}\bar{h}_{00}\delta_{ij}
\label{h in terms of trace-reversed h}%
\end{equation}
Using $A^{2}\equiv A^{i}A^{i}$, it is found that$\left(  \ref{H}\right)  $
simplifies to%
\begin{align}
H  &  =mc^{2}+\dfrac{1}{2m}\left(  P^{i}-eA^{i}\right)  ^{2}+e\varphi
-\dfrac{1}{4}\bar{h}_{00}mc^{2}+\dfrac{1}{8m}\bar{h}_{00}P^{2}-c\bar{h}%
_{0i}P^{i}\nonumber\\
& \nonumber\\
&  +\dfrac{e}{2}\bar{h}_{00}\varphi+\dfrac{5e^{2}}{8m}\bar{h}_{00}A^{2}%
+\dfrac{e}{2m}\left(  \dfrac{1}{c}\varphi\bar{h}_{0i}-\dfrac{3}{2}\bar{h}%
_{00}A^{i}\right)  P^{i}%
\end{align}
It is evident that even in this approximation of non-relativistic
gravitational sources, there are still additional terms in the Hamiltonian
that are missing in $\left(  \ref{H_DeWitt '}\right)  $. Furthermore, notice
that DeWitt combines the terms involving $P^{i}$, $A^{i}$, and $h_{0i}$ into a
single term expressed as $\left(  P^{i}-eA^{i}-cmh_{0i}\right)  ^{2}$. This
leads to the following errors.

\begin{itemize}
\item It implies a minimal coupling rule given by $\vec{P}\rightarrow\vec
{P}-e\vec{A}-m\vec{h}$ rather than the full minimal coupling rule found in
$\left(  \ref{minimal coupling}\right)  $.

\item It predicts a coupling term of the form $\frac{1}{2m}qA^{i}h_{0i}$ in
the Hamiltonian which is absent when the electromagnetic field in the
Lagrangian is properly expressed as $g_{\mu\nu}A^{\mu}v^{\nu}$ instead of
$A_{\mu}v^{\mu}$.

\item It implies a term involving $m\vec{h}^{2}$ which is not consistent with
a linear approximation.\footnote{The minimal coupling rule from DeWitt's
formulation, and the associated couplings in the Hamiltonian, $qA^{i}h_{0i}$
and $m\vec{h}^{2}$, are described, respectively, in \cite{Gravitational AB} as
predicting an interaction between EM\ and GR radiation\ fields mediated by the
quantum system, and a gravitational Landau-diamagnetism type of interaction of
the quantum system with GR\ radiation.
\par
{}}\vspace{0.3in}\vspace{0.3in}
\end{itemize}

\noindent\underline{\textbf{II. London equations, Meissner effects, and
penetration depths}}\bigskip

In this section, gravito-electromagnetic field equations are introduced in
harmonic coordinates, then constitutive equations are developed from the
canonical momentum and combined with the field equations to obtain new London
equations. Using the trace-reversed metric perturbation, $\bar{h}_{\mu\nu
}\equiv h_{\mu\nu}-\tfrac{1}{2}\eta_{\mu\nu}h$, and the harmonic coordinate
condition, $\partial^{\nu}\bar{h}_{\mu\nu}=0$, makes the linearized Einstein
equation, $G^{\mu\nu}=\kappa T^{\mu\nu}$, become $\square\bar{h}^{\mu\nu
}=-2\kappa T^{\mu\nu}$, where $\kappa=8\pi G/c^{4}$. For a non-relativistic
ideal fluid, we have%
\begin{equation}
T^{00}=\rho c^{2},\qquad T^{0i}=\rho cV^{i},\qquad T^{ij}=0
\end{equation}
which means $\bar{h}^{ij}=0$. A gravito-scalar potential and gravito-vector
potential can be defined respectively as%
\begin{equation}
\varphi_{G}\equiv-\dfrac{c^{2}}{4}\bar{h}_{00}=-\dfrac{c^{2}}{4}\bar{h}%
^{00}\qquad\text{and}\qquad h^{i}\equiv\dfrac{c}{4}\bar{h}_{0i}=-\dfrac{c}%
{4}\bar{h}^{0i}\label{phi_g and h_vector}%
\end{equation}
A gravito-electric field (the Newtonian gravitational field) and a
gravito-magnetic field (Lense-Thirring field) can also be defined respectively
as\footnote{Note that the harmonic coordinate condition, $\partial_{\nu}%
\bar{h}^{\mu\nu}=0$, leads to $\partial_{0}\bar{h}^{i0}+\partial_{j}\bar
{h}^{ij}=0$. Since non-relativistic sources led to $\bar{h}^{ij}=0$, then
$\partial_{t}\bar{h}^{i0}=0$ which means that $\vec{h}$ is time-independent in
this approximation. For a discussion of this topic, see \cite{Mirages}.}%
\begin{equation}
\vec{E}_{G}\equiv-\nabla\varphi_{G}\text{\qquad and\qquad}\vec{B}_{G}%
\equiv\nabla\times\vec{h}\label{E_G and B_G}%
\end{equation}
Defining the mass current density as $J_{m}^{i}=T^{0i}/c=\rho_{m}V_{i}$, where
$\rho_{m}$ is mass density, leads to non-homogeneous field equations given by%
\begin{equation}
\nabla\cdot\vec{E}_{G}=-\rho_{m}/\varepsilon_{G}\text{\qquad and\qquad}%
\nabla\times\vec{B}_{G}=-\mu_{G}\vec{J}_{m}+\tfrac{1}{c^{2}}\partial_{t}%
\vec{E}_{G}\label{Gauss and Ampere}%
\end{equation}
where $\varepsilon_{G}\equiv\left(  4\pi G\right)  ^{-1}$ and $\mu_{G}%
\equiv4\pi G/c^{2}$. These can be described as a gravito-Gauss law (Newton's
law of gravitation), and a gravito-Ampere law, respectively.\vspace{0.3in}

To develop constitutive equations for a superconductor, we begin by promoting
the canonical momentum in $\left(  \ref{canonical momentum}\right)  $ to a
quantum mechanical operator, $\hat{P}_{i}=-i\hslash\partial_{i}$, and act on
the complex order parameter, $\Psi\left(  r\right)  =\psi\left(  r\right)
e^{i\vec{p}\cdot\vec{r}}$, where $\psi^{2}=n_{s}$ is the number density of
Cooper pairs. This gives\footnote{This can be considered a semiclassical
approach where the gravitational field, $h_{\mu\nu}$, is still a
\textit{classical} field while $\hat{p}$, $\hat{v}$, and $\hat{A}$ are quantum
operators that act on the Cooper pair state, $\Psi$.
\par
{}}%
\begin{equation}
\hat{P}_{i}\Psi=\left[  \gamma m\left(  cg_{0i}+g_{ij}\hat{v}^{j}\right)
-q\left(  g_{0i}\hat{A}^{0}+g_{ij}\hat{A}^{j}\right)  \right]  \Psi
\end{equation}
Since the bulk of the superconductor is in the zero-momentum eigenstate, then
$\hat{P}_{i}\Psi=p_{0}\Psi=0$. Then taking the expectation value gives%
\begin{equation}
0=\gamma m\left(  cg_{0i}+g_{ij}\left\langle \hat{v}_{i}\right\rangle \right)
-q\left(  g_{0i}\left\langle \hat{A}^{0}\right\rangle +g_{ij}\left\langle
\hat{A}^{j}\right\rangle \right)  \label{zero momentum}%
\end{equation}
Applying Ehrenfest's theorem allows this equation to return to a classical
equation of motion once again. To first order in the metric perturbation, and
first order in test mass velocity, $\left(  \ref{gamma}\right)  $ becomes
$\gamma\approx1+h_{00}/2+h_{0j}v^{j}/c$. Then $\left(  \ref{zero momentum}%
\right)  $ becomes%
\begin{equation}
\left(  1+h_{00}/2\right)  mv_{i}=-m\left(  ch_{0i}+h_{ij}v^{j}\right)
+q\left(  A_{i}+h_{0i}A^{0}+h_{ij}A^{j}\right)
\end{equation}
Using $\left(  \ref{h in terms of trace-reversed h}\right)  $ and $\left(
\ref{phi_g and h_vector}\right)  $ leads to%
\begin{equation}
v_{i}=\frac{q}{m}\vec{A}+\frac{q}{mc^{2}}\varphi_{G}\vec{A}-4\vec{h}%
+\dfrac{4q}{mc^{2}}\varphi\vec{h} \label{v}%
\end{equation}
The charge and mass supercurrent densities are, respectively,\footnote{Note
that a negative is used in $\vec{J}_{c}=-n_{s}q\vec{v}$ so that when $q=-e$ is
used, then $\vec{J}_{c}$ becomes positive and hence represents the
\textit{conventional} current.
\par
{}}%
\begin{equation}
\vec{J}_{c}=-n_{s}q\vec{v}\qquad\text{and}\qquad\vec{J}_{m}=n_{s}m\vec{v}
\label{Mass and charge currents}%
\end{equation}
where $n_{s}$ is the number density of Cooper pairs. Inserting $\left(
\ref{v}\right)  $ into $\left(  \ref{Mass and charge currents}\right)  $, and
using $q=-e$ and $m=m_{e}$ for electrons, leads to%
\begin{equation}
\vec{J}_{c}=-\Lambda_{L}\left(  \alpha\vec{A}-\beta\vec{h}\right)  \label{J_c}%
\end{equation}
and%
\begin{equation}
\vec{J}_{m}=n_{s}e\left(  \alpha\vec{A}-\beta\vec{h}\right)  \label{J_m}%
\end{equation}
where $\Lambda_{L}\equiv n_{s}e^{2}/m_{e}$ is the London constant
\cite{London}, and\footnote{Similar expressions to $\left(  \ref{J_c}\right)
$ and $\left(  \ref{J_m}\right)  $ can be found in \cite{Li and Torr},
however, with $\alpha=1$ and $\beta=m_{e}/e$.}%
\begin{equation}
\alpha\equiv1+\varphi_{G}/c^{2}\qquad\text{and}\qquad\beta\equiv\frac{4\left(
m_{e}c^{2}+e\varphi\right)  }{ec^{2}} \label{alpha and beta}%
\end{equation}
The expressions in $\left(  \ref{J_c}\right)  $ and $\left(  \ref{J_m}\right)
$ are the London constitutive equations for a non-relativistic supercurrent in
the presence of electromagnetic fields and gravitational fields (from
non-relativistic sources). Notice that if all gravitational fields are
neglected, then $\alpha=1$ and $\vec{h}=0$, so $\left(  \ref{J_c}\right)  $
becomes the well-known London constitutive equation, $\vec{J}_{c}=-\Lambda
_{L}\vec{A}$. Also notice that for the case of a neutral superfluid, setting
the charge to zero in $\left(  \ref{J_m}\right)  $ gives $\vec{J}_{m}%
=-n_{s}m_{e}\vec{h}$ which is the constitutive equation for a neutral
superfluid in the presence of a gravito-vector potential.\vspace{0.3in}

Taking the time-derivative of $\left(  \ref{J_c}\right)  $ and $\left(
\ref{J_m}\right)  $, and using the fact that $\nabla\varphi=0$ inside a
superconductor\footnote{In standard London theory, the requirement that
$\nabla\varphi=0$ inside a superconductor follows from inserting $\vec
{E}=-\nabla\varphi-\partial_{t}\vec{A}$ into the electric London equation,
$\partial_{t}\vec{J}_{c}=\Lambda_{L}\vec{E}$ which gives $\partial_{t}\left(
\vec{J}_{c}+\Lambda_{L}\vec{A}\right)  =-\Lambda_{L}\nabla\varphi$. Since
$\vec{J}_{s}=-\Lambda_{L}\vec{A}$ is the London constitutive equation, then it
follows that $\nabla\varphi=0$. However, this assumption is not taken for
granted and is widely discussed in the literature, as summarized in
\cite{Lipavsky et al}. In the treatment used here, we can also insert $\vec
{E}=-\nabla\varphi-\partial_{t}\vec{A}$ into $\left(
\ref{time-derivative of J_c}\right)  $ to obtain $\partial_{t}\left[  \vec
{J}_{c}+\Lambda_{L}\left(  \alpha\vec{A}-\beta\vec{h}\right)  \right]
=-\Lambda_{L}\alpha\nabla\varphi$. Then using $\left(  \ref{J_c}\right)  $
requires the bracket is zero and therefore $\nabla\varphi=0$.}, and
$\partial_{t}\vec{h}=0$ in this approximation, leads to%
\begin{equation}
\partial_{t}\vec{J}_{c}=\Lambda_{L}\left(  \alpha\vec{E}-\frac{1}{c^{2}}%
\dot{\varphi}_{G}\vec{A}-\frac{4}{c^{2}}\dot{\varphi}\vec{h}\right)
\label{time-derivative of J_c}%
\end{equation}
and%
\begin{equation}
\partial_{t}\vec{J}_{m}=-en_{s}\left(  \alpha\vec{E}-\frac{1}{c^{2}}%
\dot{\varphi}_{G}\vec{A}-\frac{4}{c^{2}}\dot{\varphi}\vec{h}\right)
\label{time-derivative of J_m}%
\end{equation}
Notice that $\left(  \ref{time-derivative of J_c}\right)  $ is the usual
electric London equation, $\partial_{t}\vec{J}_{c}=\Lambda_{L}\vec{E}$, but
with correction terms due to gravity. However, $\left(
\ref{time-derivative of J_c}\right)  $ is a redundant equation since $\vec
{J}_{c}=-\vec{J}_{m}\left(  e/m_{e}\right)  $.\vspace{0.3in}

Taking the curl of $\left(  \ref{J_c}\right)  $ and $\left(  \ref{J_m}\right)
$ leads to%
\begin{equation}
\nabla\times\vec{J}_{c}=-\Lambda_{L}\left(  \alpha\vec{B}-\beta\vec{B}%
_{G}-\frac{1}{c^{2}}\vec{E}_{G}\times\vec{A}\right)  \label{curl of J_c}%
\end{equation}
and%
\begin{equation}
\nabla\times\vec{J}_{m}=n_{s}e\left(  \alpha\vec{B}-\beta\vec{B}_{G}-\frac
{1}{c^{2}}\vec{E}_{G}\times\vec{A}\right)  \label{curl of J_m}%
\end{equation}
Notice that $\left(  \ref{curl of J_c}\right)  $ is the usual magnetic London
equation, $\nabla\times\vec{J}_{c}=-\Lambda_{L}\vec{B}$, but with correction
terms due to gravity. However, for the case of a neutral superfluid, setting
the charge to zero makes $\left(  \ref{curl of J_m}\right)  $ become
$\nabla\times\vec{J}_{m}=-4n_{s}m_{e}\vec{B}_{G}$. This can be viewed as a
gravito-magnetic London equation for a neutral superfluid.\vspace{0.3in}

Note that the usual London gauge is $\nabla\cdot\vec{A}\propto\nabla\cdot
\vec{J}_{c}=0$. By the continuity equation, this means $\partial_{t}\rho
_{c}=0$ which is consistent with a static number density of Cooper pairs since
$\rho_{c}=2n_{s}e$. Applying the same condition to the modified London
constitutive equation in $\left(  \ref{J_c}\right)  $ leads to a new London
condition given by $\nabla\cdot\left(  \alpha\vec{A}-\beta\vec{h}\right)  =0$.
Using $\left(  \ref{alpha and beta}\right)  $ and $\nabla\varphi=0$ leads to%
\begin{equation}
\alpha\nabla\cdot\vec{A}-\frac{1}{c^{2}}\left(  \vec{E}_{G}\cdot\vec
{A}\right)  -\beta\nabla\cdot\vec{h}=0 \label{London gauge}%
\end{equation}
This is the modified London gauge condition associated with the London
equations developed above. Since we are using harmonic coordinates,
$\partial_{\nu}\bar{h}^{\mu\nu}=0$, then the last term in $\left(
\ref{London gauge}\right)  $ can also be expressed using $\nabla\cdot\vec
{h}=-\dot{\varphi}_{G}/c^{2}$.\pagebreak

For simplicity, consider a cylindrically symmetric superconductor with an axis
normal to the surface of Earth so that $\vec{A}$ is azimuthal and $\vec{E}%
_{G}$ (due to earth) is along the axis of the cylinder which means $\vec
{A}\times\vec{E}_{G}=0$. For a steady-state supercurrent, Ampere's law is
$\nabla\times\vec{B}=\mu_{0}\vec{J}_{c}$, and the gravito-Ampere law is
$\nabla\times\vec{B}_{G}=-\mu_{G}\vec{J}_{m}$. Taking the curl of both field
equations, using the identity $\nabla\times\nabla\times\vec{B}=\nabla\left(
\nabla\cdot\vec{B}\right)  -\nabla^{2}\vec{B}$, where $\nabla\cdot\vec{B}=0$,
and using $\left(  \ref{curl of J_c}\right)  $ and $\left(  \ref{curl of J_m}%
\right)  $ in their respective field equations, leads to\footnote{Coupled
differential equations similar to $\left(  \ref{1}\right)  $ and $\left(
\ref{2}\right)  $ can also be found in \cite{N. Li} and \cite{Peng (new)}.}%
\begin{equation}
\nabla^{2}\vec{B}=\mu_{0}\Lambda_{L}\left(  \alpha\vec{B}-\beta\vec{B}%
_{G}\right)  \label{1}%
\end{equation}
and%
\begin{equation}
\nabla^{2}\vec{B}_{G}=\mu_{G}n_{s}e\left(  \alpha\vec{B}-\beta\vec{B}%
_{G}\right)  \label{2}%
\end{equation}
Notice that neglecting gravity makes $\alpha=1$ and $\vec{B}_{G}=0$. Then
$\left(  \ref{1}\right)  $ becomes a Yukawa-like equation, $\nabla^{2}\vec
{B}=\mu_{0}\Lambda_{L}\vec{B}$, where $\lambda_{L}=1/\sqrt{\mu_{0}\Lambda_{L}%
}$ is the London penetration depth. The only non-trivial physical solution is
$B\left(  z\right)  =B_{0}e^{-z/\lambda_{L}}$, where $z$ is the distance from
the surface to the interior of the superconductor, and $B_{0}$ is the
magnitude of the magnetic field at the surface of the superconductor. This is
the standard Meissner effect.\vspace{0.3in}

Also notice that for a neutral superfluid, $\left(  \ref{2}\right)  $ becomes
$\nabla^{2}\vec{B}_{G}=-4\mu_{G}n_{s}m_{e}\vec{B}_{G}$ which is a
Helmholtz-like differential equation (rather than a Yukawa-like differential
equation), and therefore only allows \textit{sinusoidal} solutions, not
\textit{exponential} solutions. Since there is no exponential decay of the
field then there is no penetration depth and no associated Meissner effect.
The reason can be traced back to the difference in the sign appearing in the
magnetic field equation, $\nabla\times\vec{B}=\mu_{0}\vec{J}_{c}$, and the
gravito-magnetic field equation, $\nabla\times\vec{B}_{G}=-\mu_{G}\vec{J}_{m}%
$. The negative sign in the gravito-Ampere law eliminates a gravitational
Meissner effect for a neutral superfluid. Physically speaking, this implies a
paramagnetic effect instead of a diamagnetic (Meissner)\textit{\ }effect. This
is in agreement with \cite{Ciubotariu}, \cite{Agop} but in disagreement with
\cite{Peng (new)}.\vspace{0.3in}

In fact, for a maximum gravito-magnetic field at the surface of the
superconductor, the solution to the gravito-Ampere field equation would have
the form $B_{G}=B_{G,0}\cos\left(  kz\right)  $, where $k^{2}=4\mu_{G}%
n_{s}m_{e}$. The associated spatial periodicity in the field would be given by
$2\pi/k$ which gives%
\begin{equation}
\lambda_{\left(  \text{periodicity of }\vec{B}_{G}\right)  }\equiv\frac{2\pi
}{\sqrt{4\mu_{G}n_{s}m_{e}}} \label{lambda_(per)}%
\end{equation}
To obtain a numerical estimate for this quantity, consider if each atom
contributes two conduction electrons, and only $10^{-3}$ of the conduction
electrons are in a superconducting state, then $n_{s}\approx2n\left(
10^{-3}\right)  $, where $n=\rho_{m}/m$ is the number density of atoms. For
Niobium, the mass density is $\rho_{m}\approx8.6\times10^{3}$kg/m$^{\text{3}}$
and the mass per atom is $m\approx1.5\times10^{-25}$kg/atom. Then the number
density of atoms is $n\approx5.7\times10^{28}$m$^{-3}$ and therefore the
number density of Cooper pairs is $n_{s}\approx2n\left(  10^{-3}\right)
\approx1.1\times10^{26}$m$^{-3}$. Then $\left(  \ref{lambda_(per)}\right)  $
gives approximately $3.3\times10^{15}$m which is clearly not measurable on a
terrestrial scale.\vspace{0.3in}

For a charged supercurrent in the presence of \textit{both} magnetic and
gravito-magnetic fields, the differential equations $\left(  \ref{1}\right)  $
and $\left(  \ref{2}\right)  $ need to be decoupled to obtain solutions.
Solving $\left(  \ref{1}\right)  $ for $\vec{B}_{G}$, substituting the result
into $\left(  \ref{2}\right)  $, and canceling common terms gives%
\begin{equation}
\nabla^{2}\left[  \nabla^{2}\vec{B}-\lambda_{L\left(  \text{modified}\right)
}^{-2}\vec{B}\right]  =0 \label{Double laplacian for B}%
\end{equation}
where a \textquotedblleft modified\textquotedblright\ London penetration due
to the presence of gravity is defined as%
\begin{equation}
\lambda_{L\left(  \text{modified}\right)  }^{-2}\equiv\alpha\lambda_{L}%
^{-2}-\beta\mu_{G}n_{s}e \label{Lamba_L (mod)}%
\end{equation}
Assuming $\nabla^{2}\vec{B}$ and $\vec{B}$ both go to zero as $r\rightarrow
\infty$, then the solution to $\left(  \ref{Double laplacian for B}\right)  $
requires that the bracket is zero which leads to $\nabla^{2}\vec{B}%
=\lambda_{L\left(  \text{mod}\right)  }^{-2}\vec{B}$. Notice that the first
term in $\left(  \ref{Lamba_L (mod)}\right)  $ encodes a correction due to the
gravitational scalar potential since $\alpha\equiv1+\varphi_{G}/c^{2}$. The
second term in $\left(  \ref{Lamba_L (mod)}\right)  $ encodes a correction due
to the gravito-magnetic field since it can be traced back to the terms
involving $\beta\vec{B}_{G}$ in $\left(  \ref{1}\right)  $ and $\left(
\ref{2}\right)  $.\vspace{0.3in}

An order of magnitude can be calculated for each term in $\left(
\ref{Lamba_L (mod)}\right)  $. For example, the London penetration depth for
niobium is known to be $\lambda_{L}\sim10^{-9}$ m. At the surface of earth,
$\varphi_{G}/c^{2}\sim10^{-10}$ which means the correction to the London
penetration due to the earth's gravitational scalar potential is on the order
of $10^{-19}$ m and therefore not observable. For the second term in $\left(
\ref{Lamba_L (mod)}\right)  $, we can note from $\left(  \ref{alpha and beta}%
\right)  $ that if $m_{e}c^{2}>>e\varphi$, then $\beta\approx4m_{e}%
/e\sim10^{-11}$kg/C. Also using $n_{s}\sim10^{26}$m$^{-3}$ means the second
term in $\left(  \ref{Lamba_L (mod)}\right)  $ is $\sim10^{-30}$m$^{-2}$.
Since the first term is $\alpha\lambda_{L}^{-2}\sim10^{18}$ m$^{-2}$, then the
second term is completely negligible. Hence we find that the presence of a
Newtonian and/or Lense-Thirring gravitational field cannot have a measurable
effect on the penetration depth of the magnetic field.\vspace{0.3in}

Next, solving $\left(  \ref{2}\right)  $ for $\vec{B}$, substituting the
result into $\left(  \ref{1}\right)  $, and canceling common terms gives%
\begin{equation}
\nabla^{2}\left[  \nabla^{2}\vec{B}_{G}+k_{\left(  \text{modified}\right)
}^{2}\vec{B}_{G}\right]  =0
\end{equation}
where%
\begin{equation}
k_{\left(  \text{modified}\right)  }^{2}\equiv\mu_{G}n_{s}e\beta-\alpha\mu
_{0}\Lambda_{L} \label{k_(mod)}%
\end{equation}
Assuming $\nabla^{2}\vec{B}$ and $\vec{B}$ both go to zero as $r\rightarrow
\infty$, then the solution to $\left(  \ref{Double laplacian for B}\right)  $
requires that the bracket is zero which leads to $\nabla^{2}\vec{B}%
_{G}=k_{\left(  \text{mod}\right)  }^{2}\vec{B}$. Notice that for a neutral
superfluid (or in the absence of a magnetic field), the expression in $\left(
\ref{k_(mod)}\right)  $ reduces to $k^{2}=4\mu_{G}n_{s}m_{e}$ which leads to a
paramagnetic effect, as stated before.\vspace{0.3in}

However, for a charged supercurrent, the first term in $\left(  \ref{k_(mod)}%
\right)  $ encodes a correction due to the electric scalar potential, while
the second term encodes a correction due to the gravitational scalar potential
and the electric charge. An order of magnitude can be calculated for the first
term using $\beta\approx4m_{e}/e\sim10^{-11}$kg/C and $n_{s}\sim10^{26}%
$m$^{-3}$ which gives $\sim10^{-30}$m$^{-2}$. Since $\mu_{0}\Lambda
_{L}=\lambda_{L}^{-2}$, then the second term can be expressed in terms of the
London penetration depth which gives $\alpha\lambda_{L}^{-2}\sim10^{18}$
m$^{-2}$. In that case, the second term in $\left(  \ref{k_(mod)}\right)  $
far exceeds the first term, and therefore $\left(  \ref{k_(mod)}\right)  $ is
negative. This leads to a diamagnetic (Meissner)\textit{\ }effect. In fact,
the gravito-magnetic field is expelled with effectively the same London
penetration depth as the magnetic field. The possibility of a gravito-magnetic
Meissner effect is in agreement with \cite{ABC}, \cite{Agop (2)}, \cite{Agop
(3)}.\vspace{0.3in}

It is helpful to compare the physics contained in $\left(  \ref{Lamba_L (mod)}%
\right)  $ and $\left(  \ref{k_(mod)}\right)  $. They are analogous in the
sense that $\left(  \ref{Lamba_L (mod)}\right)  $ predicts a diamagnetic
(Meissner) effect for the magnetic field, but it is altered by the presence of
the gravito-magnetic field, while $\left(  \ref{k_(mod)}\right)  $ predicts a
paramagnetic effect for the gravito-magnetic field but it is altered by the
presence of the magnetic field \ In the case of $\left(  \ref{Lamba_L (mod)}%
\right)  $, the alteration due to the presence of the gravito-magnetic field
is extremely small, so that the diamagnetic (Meissner) effect remains but with
a slightly larger penetration depth. However, for the case of $\left(
\ref{k_(mod)}\right)  $, the alteration due to the presence of the magnetic
field is so substantial that it switches a paramagnetic effect into a
diamagnetic (Meissner) effect for the gravito-magnetic field.\bigskip

Hence the findings above are summarized as follows:

\begin{itemize}
\item Supercurrent in the presence of only $\vec{B}$: magnetic Meissner effect.

\item Supercurrent in the presence of only $\vec{B}_{G}$: no gravito-magnetic
Meissner effect.

\item Supercurrent in the presence of \textit{both} $\vec{B}$ and $\vec{B}%
_{G}$: both magnetic and gravito-magnetic Meissner effects.

\item Neutral superfluid in the presence of $\vec{B}_{G}$: no gravito-magnetic
Meissner effect.\footnote{A neutral superfluid in the presence of $\vec{B}$ is
not expected to have any interaction simply because there is no charge to
couple to the magnetic field.}\vspace{0.3in}
\end{itemize}

These results demonstrate an important interaction between electromagnetism,
gravitation, and a quantum mechanical system that only occurs when all three
are present. The superconductor provides the quantum mechanical system which
is necessary to have any kind of Meissner effect. The gravito-magnetic field
is required to create a novel gravitational effect. Lastly, the magnetic field
is necessary to mediate the interaction. In the absence of a magnetic field,
the novel gravitational effect would not take place.\bigskip

A final important consideration is the issue of coordinate-freedom in
linearized General Relativity. The gravito-magnetic field, $\vec{B}_{G}%
=\nabla\times\vec{h}$, is a coordinate-dependent quantity which can be made to
vanish by a linear coordinate transformation, $x^{\prime\mu}=x^{\mu}-\xi^{\mu
}$. Since the linearized metric perturbation transforms as%
\begin{equation}
h_{\mu\nu}^{\prime}=h_{\mu\nu}+\partial_{\mu}\xi_{\nu}+\partial_{\nu}\xi_{\mu}%
\end{equation}
then $\vec{B}_{G}$ transforms as%
\begin{equation}
\vec{B}_{G}^{\prime}=\vec{B}_{G}-\tfrac{1}{4}\nabla\times\overset{_{\cdot
}}{\vec{\xi}}%
\end{equation}
Therefore, the effects associated with $\vec{B}_{G}$ can be made to vanish.
Alternatively, a coordinate-invariant approach can be used which also applies
to gravitational waves. This is discussed in \cite{dissertation},
\cite{FQMT15}, \cite{Essay}.\pagebreak

\noindent\underline{\textbf{III. Flux quantum in the body of a superconductor
}}\bigskip

In Ginzburg-Landau theory, the minimal coupling rule, $\hat{P}_{i}%
\rightarrow\hat{P}_{i}-q\hat{A}_{i}$, makes the supercurrent become
\cite{Tinkham}%
\begin{equation}
\vec{J}=\dfrac{e}{2m}\left[  \Psi^{\ast}\left(  -i\hslash\nabla\right)
\Psi-\Psi\left(  -i\hslash\nabla\right)  \Psi^{\ast}-2e\vec{A}\left\vert
\Psi\right\vert ^{2}\right]  \label{J}%
\end{equation}
where $\Psi\left(  r\right)  $ is the complex order parameter. Using $\left(
\ref{h in terms of trace-reversed h}\right)  $ and $\left(
\ref{phi_g and h_vector}\right)  $ in$\ \left(  \ref{minimal coupling}\right)
$, and promoting the canonical momentum to a quantum mechanical operator makes
the minimal coupling rule become%
\begin{equation}
\hat{P}_{i}\quad\rightarrow\quad\hat{P}_{i}-\gamma m\left(  4h^{i}-\frac
{2}{c^{2}}\varphi_{G}v_{i}\right)  +q\left(  A_{i}+\frac{4}{c^{2}}\varphi
h^{i}-\frac{2}{c^{2}}\varphi_{G}A_{i}\right)  \label{minimal coupling '}%
\end{equation}
To first order in the metric perturbation, and first order in test mass
velocity, we can use $\gamma\approx1$ in $\left(  \ref{minimal coupling '}%
\right)  $. For convenience, we can also define the entire coupling vector as%
\begin{equation}
\vec{C}\equiv-m\left(  4\vec{h}-\frac{2}{c^{2}}\varphi_{G}\vec{v}\right)
+q\left(  \vec{A}+\frac{4}{c^{2}}\varphi\vec{h}-\frac{2}{c^{2}}\varphi_{G}%
\vec{A}\right)  \label{C}%
\end{equation}
Then the corresponding supercurrent becomes%
\begin{equation}
\vec{J}=\dfrac{e}{2m}\left[  \Psi^{\ast}\left(  -i\hslash\nabla\right)
\Psi-\Psi\left(  -i\hslash\nabla\right)  \Psi^{\ast}+\vec{C}\left\vert
\Psi\right\vert ^{2}\right]
\end{equation}
which reduces to $\left(  \ref{J}\right)  $ in the absence of gravitation.
Using $\Psi\left(  r\right)  =\sqrt{n_{s}\left(  r\right)  }e^{i\theta\left(
r\right)  }$, where $\theta$ is the phase, leads to%
\begin{equation}
\vec{J}=\dfrac{e}{2m}\left(  2\hslash\nabla\theta+\vec{C}\right)  n_{s}
\label{J_vector}%
\end{equation}
In the previous section, it was shown that inside the body of the
superconductor (beyond the London penetration depth), all the fields in
$\left(  \ref{v}\right)  $ vanish and therefore the supercurrent velocity is
zero. Therefore, using $J_{i}=0$ in $\left(  \ref{J_vector}\right)  $ and
$v_{i}=0$ in $\left(  \ref{C}\right)  $ makes $\left(  \ref{J_vector}\right)
$ become\footnote{An expression similar to $\left(  \ref{J_vector}\right)  $
is found in \cite{N. Li}, however, the terms involving $\varphi$ and
$\varphi_{G}$ in $\left(  \ref{C}\right)  $ are missing.}%
\begin{equation}
\hslash\nabla\theta=4m\vec{h}-q\vec{A}-\frac{4q}{c^{2}}\varphi\vec{h}%
+\frac{2q}{c^{2}}\varphi_{G}\vec{A}%
\end{equation}
Integrating around a closed loop gives%
\begin{equation}
\hslash\oint\limits_{C}\left(  \nabla\theta\right)  \cdot d\vec{l}%
=\oint\limits_{C}\left(  4m\vec{h}-q\vec{A}-\frac{4q}{c^{2}}\varphi\vec
{h}+\frac{2q}{c^{2}}\varphi_{G}\vec{A}\right)  \cdot d\vec{l}%
\end{equation}
Since the order parameter is single-valued, then it must return to the same
value when the line integral returns to the same point. Therefore, the left
side must be $2\pi n$, where $n$ is an integer. Applying Stokes' theorem on
the right side gives%
\begin{equation}
\hslash2\pi n=\int\limits_{S}\nabla\times\left(  4m\vec{h}-q\vec{A}-\frac
{4q}{c^{2}}\varphi\vec{h}+\frac{2q}{c^{2}}\varphi_{G}\vec{A}\right)  \cdot
d\vec{S}%
\end{equation}
where $S$ is the surface bounded by $C$. Using $\nabla\varphi=0$ within the
superconductor, and $q=-2e$ and $m=2m_{e}$ for Cooper pairs, gives%
\begin{equation}
e\left(  1-\frac{2\varphi_{G}}{c^{2}}\right)  \Phi_{B}+4m_{e}\left(
1-\frac{q}{2m_{e}c^{2}}\varphi\right)  \Phi_{B_{G}}+\frac{2e}{c^{2}}%
\int\limits_{S}\left(  \vec{E}_{G}\times\vec{A}\right)  \cdot d\vec{S}%
=n\frac{h}{2}%
\end{equation}
where $\Phi_{B}$ and $\Phi_{B_{G}}$ are the flux of $\vec{B}$ and $\vec{B}%
_{G}$, respectively. Since $\varphi_{G}<<c^{2}$ and $e\varphi<<m_{e}c^{2}$,
then the result can be approximated to%
\begin{equation}
e\Phi_{B}+4m_{e}\Phi_{B_{G}}+\frac{2e}{c^{2}}\int\limits_{S}\left(  \vec
{E}_{G}\times\vec{A}\right)  \cdot d\vec{S}=n\frac{h}{2}
\label{quantized flux}%
\end{equation}
This result is consistent with \cite{Peng (new)}, \cite{Ross}, and DeWitt's
statement in \cite{DeWitt} that \textquotedblleft the total flux of $\vec{G}$
linking a superconducting circuit must be quantized in units of $\frac{1}{2}%
h$,\textquotedblright\ where $\vec{G}=e\nabla\times\vec{A}+m\nabla\times
\vec{h}$. However, none of these authors have the additional term in $\left(
\ref{quantized flux}\right)  $ involving the flux of $\vec{E}_{G}\times\vec
{A}$.\vspace{0.3in}\vspace{0.3in}

\noindent\underline{\textbf{IV. Quantized supercurrent for a superconducting
ring in the presence of a \textquotedblleft mass solenoid\textquotedblright}%
}\bigskip

To obtain the quantized supercurrent, we begin by promoting the canonical
momentum in $\left(  \ref{canonical momentum}\right)  $ to a quantum
mechanical operator and act on the complex order parameter.%
\begin{equation}
\hat{P}_{i}\Psi=\gamma m\left(  cg_{0i}+g_{ij}\hat{v}^{j}\right)  -q\left(
g_{0i}\hat{A}^{0}+g_{ij}\hat{A}^{j}\right)  \Psi\label{canonical momentum ''}%
\end{equation}
Again we use $\Psi=\psi e^{i\theta}$, but now let $\psi=\sqrt{n_{s}}$ be a
uniform number density around a ring. Therefore, $\hat{P}_{i}\Psi=\hslash
\Psi\partial^{i}\theta$. Then using $\left(  \ref{canonical momentum ''}%
\right)  $ and taking the expectation value gives%
\begin{equation}
\hslash\left\langle \partial^{i}\theta\right\rangle =\gamma m\left(
cg_{0i}+g_{ij}\left\langle \hat{v}^{j}\right\rangle \right)  -q\left(
g_{0i}\left\langle \hat{A}^{0}\right\rangle +g_{ij}\left\langle \hat{A}%
^{j}\right\rangle \right)
\end{equation}
Applying Ehrenfest's theorem allows this equation to return to a classical
equation once again. To first order in the metric perturbation, and first
order in test mass velocity, we can use $\gamma\approx1+h_{00}/2$. Also using
$\left(  \ref{h in terms of trace-reversed h}\right)  $ and $\left(
\ref{phi_g and h_vector}\right)  $gives%
\begin{equation}
\hslash\nabla\theta=\left(  1-\frac{3\varphi_{G}}{c^{2}}\right)  m\vec
{v}-\left(  1-\frac{2\varphi_{G}}{c^{2}}\right)  q\vec{A}+4\left(
1-\frac{q\varphi}{mc^{2}}\right)  m\vec{h}%
\end{equation}
We can write this expression in terms of the supercurrent density which is
$J^{i}=-qn_{s}v^{i}$ and integrate around a circular superconducting ring.%
\begin{equation}
\hslash\oint\limits_{C}\left(  \nabla\theta\right)  \cdot d\vec{l}%
=\oint\limits_{C}\left[  -\frac{m}{qn_{s}}\left(  1-\frac{3\varphi_{G}}{c^{2}%
}\right)  \vec{J}-\left(  1-\frac{2\varphi_{G}}{c^{2}}\right)  q\vec
{A}+4\left(  1-\frac{q\varphi}{mc^{2}}\right)  m\vec{h}\right]  \cdot d\vec{l}%
\end{equation}
Since the order parameter is single-valued, then it must return to the same
value when the line integral returns to the same point. Therefore, the left
side must be $2\pi n$, where $n$ is an integer.\footnote{This can also be
identified as an extended application of the Byers-Yang theorem
\cite{Byers-Yang} which ordinarily applies only to a wave function in the
presence of a magnetic vector potential.} Applying Stokes' theorem on the
right side, using $q=-2e$ and $m=2m_{e}$ for Cooper pairs, and assuming
$\varphi_{G}<<c^{2}$ and $q\varphi<<mc^{2}$, gives%
\begin{equation}
\dfrac{m_{e}}{2en_{s}}\oint\limits_{C}\vec{J}\mathcal{\cdot}d\vec{l}+e\Phi
_{B}+4m_{e}\Phi_{B_{G}}+\frac{2e}{c^{2}}\int\limits_{S}\left(  \vec{E}%
_{G}\times\vec{A}+2\nabla\varphi\times\vec{h}\right)  \cdot d\vec{S}=n\frac
{h}{2} \label{Byers-Yang in curved space-time}%
\end{equation}
where $S$ is the surface bounded by $C$. The result is similar to $\left(
\ref{quantized flux}\right)  $ which applies to the bulk of the
superconductor. However, there is an additional term involving the
supercurrent density, and a term involving $\nabla\varphi\times\vec{h}$ since
$\vec{J}$ and $\nabla\varphi$ can be non-zero on the surface of the
ring.\vspace{0.3in}

As a practical example, we consider a superconducting ring in the presence of
a \textquotedblleft mass solenoid\textquotedblright\ as described in Appendix
B. This is effectively the same system that was considered by DeWitt
\cite{DeWitt}. He states, \textquotedblleft Now consider an experiment in
which the superconductor is a uniform circular ring surrounding a concentric,
axially symmetric, quasirigid mass. Suppose the mass, initially at rest, is
set in motion until a constant final angular velocity is reached. This
produces a Lense-Thirring field%
\begin{equation}
\vec{\nabla}\times\vec{h}_{0}=16\pi G\nabla^{-2}\vec{\nabla}\times\left(
\rho\vec{V}\right)  \label{del x h_0}%
\end{equation}
where $\rho$ and $\vec{V}$ are, respectively, the mass density and velocity
field of the rotating mass.\textquotedblright\footnote{Note that DeWitt's
$\vec{h}_{0}$ is related to $\vec{h}$ used in this paper by $\vec{h}=4\vec
{h}_{0}$.} For a steady-state current, the gravito-Ampere field equation in
$\left(  \ref{Gauss and Ampere}\right)  $ can indeed be written as $\nabla
^{2}\vec{h}=\left(  4\pi G/c^{2}\right)  \rho_{m}\vec{V}$, which matches
DeWitt's equation up to a factor of $4$.\vspace{0.3in}

DeWitt goes on to state, \textquotedblleft Suppose the rotating mass is kept
electromagnetically neutral (which means compensating for any Schiff-Barnhill
polarization which may be induced in it).\textquotedblright\ He also writes
the \textquotedblleft Schiff-Barnhill\textquotedblright\ field as $\vec
{F}=e\vec{E}+\frac{1}{2}m\nabla h_{00}$. Therefore, by eliminating this field,
we would be effectively eliminate $\vec{E}_{G}\times\vec{A}$ and\ $\nabla
\varphi\times\vec{h}$ found in $\left(  \ref{Byers-Yang in curved space-time}%
\right)  $ and obtain%
\begin{equation}
\dfrac{m_{e}}{2en_{s}}\oint\limits_{C}\vec{J}\mathcal{\cdot}d\vec{l}+e\Phi
_{B}+4m_{e}\Phi_{B_{G}}=n\frac{h}{2} \label{Byers-Yang in curved space-time '}%
\end{equation}
DeWitt also states, \textquotedblleft Because of the flux quantization
condition, the flux of $G$ through the superconducting ring must remain zero.
But since $\vec{\nabla}\times\vec{h}_{0}$ is nonvanishing in the final state,
a magnetic field must be induced.\textquotedblright\ This implies that DeWitt
is effectively setting $n=0$ in $\left(
\ref{Byers-Yang in curved space-time '}\right)  $. Next, DeWitt states,
\textquotedblleft Then the magnetic field must arise from a current induced in
the ring.\textquotedblright\ It is at this point that we must disagree with
DeWitt. It appears that he has also set the first term in $\left(
\ref{Byers-Yang in curved space-time '}\right)  $ \textit{independently} to
zero, then solved for $\Phi_{B}$ to obtain $\Phi_{B}=-4m_{e}\Phi_{B_{G}}/e$.
This is evident since using the definition of self-inductance, $L=\Phi_{B}/I$,
and the condition $\Phi_{B}=-4m_{e}\Phi_{B_{G}}/e$, leads to%
\begin{equation}
I=-\frac{4m_{e}}{eL}\int\limits_{S}\left(  \vec{\nabla}\times\vec{h}\right)
\cdot d\vec{S}=-\frac{16\pi Gm_{e}}{eLc^{2}}\int\limits_{S}\left(  \nabla
^{-2}\rho\vec{V}\right)  \cdot d\vec{r} \label{current}%
\end{equation}
which matches DeWitt's equation (11). However, there is a contradiction since
using the condition $\Phi_{B}=-4m_{e}\Phi_{B_{G}}/e$ required setting the
first term in $\left(  \ref{Byers-Yang in curved space-time '}\right)  $ to
zero and therefore excluded any current in the superconducting ring.\vspace
{0.3in}

Therefore, we would suggest that a correct interpretation of $\left(
\ref{Byers-Yang in curved space-time '}\right)  $ requires recognizing that
$\Phi_{B}$ and $\Phi_{B_{G}}$ are the flux of \textit{external} fields that
are introduced to the superconductor \textit{independently}. There is no
inductive relationship between them. In fact, if there is initially no
external magnetic field, then starting back at the canonical momentum in
$\left(  \ref{canonical momentum ''}\right)  $, we must set $A^{i}=0$ and
hence $\Phi_{B}$ would be absent in $\left(
\ref{Byers-Yang in curved space-time '}\right)  $ and the expression simply
becomes%
\begin{equation}
\dfrac{m_{e}}{2en_{s}}\oint\limits_{C}\vec{J}\mathcal{\cdot}d\vec{l}%
+4m_{e}\Phi_{B_{G}}=n\frac{h}{2}
\label{Byers-Yang in curved space-time ' (no B)}%
\end{equation}
If there is no persistent current initially set up in the superconducting
ring, then the first term is zero and $\left(
\ref{Byers-Yang in curved space-time ' (no B)}\right)  $ simply predicts that
the flux of any external gravito-magnetic field through the ring must be
quantized in units of $h/2$.\vspace{0.3in}

Furthermore, DeWitt obtains an order-of-magnitude estimate for his predicted
electric current\footnote{Papini \cite{Papini (GR test)} calculates a similar
result using the gravito-magnetic field of the earth which leads to an
electric current given by $I=\frac{8\pi}{5}\frac{MG}{c^{r}R}\frac{ma^{2}%
\omega}{eL}$, where $M$, $R$ and $\omega$ are, respectively the mass, radius
and angular velocity of the earth, and $L$ and $a$ are the self-inductance and
radius of the loop.} by applying Stokes' theorem to $\left(  \ref{current}%
\right)  $, integrating around the perimeter with diameter $d=2R$, and using
$\rho=M/\left(  \pi R^{2}\ell\right)  $, where $\ell$ and $M$ are the length
and mass of the cylinder, respectively, to obtain $I=-64\pi Gm_{e}M\nabla
^{-2}\vec{V}/\left(  eLc^{2}\ell d\right)  $. DeWitt sets $c=1$ and evidently
assumes $\dfrac{64\pi}{L\ell}\nabla^{-2}\sim1$ to obtain a result of%
\begin{equation}
I\sim\dfrac{Gm_{e}MV}{ed} \label{order-of-mag}%
\end{equation}
in his equation (12). However, again we must keep in mind that this result was
obtained by setting $n=0$ in $\left(  \ref{Byers-Yang in curved space-time '}%
\right)  $. Preserving this state requires that the left side remains less
than $h/2$. This means requiring $m_{e}\Phi_{B_{G}}<h/2$ and therefore
$\left(  \ref{current}\right)  $ would give $I=-\dfrac{4m_{e}}{eL}\Phi_{B_{G}%
}<\dfrac{2h}{eL}$ which is an exceedingly small electric current contrary to
DeWitt's order of magnitude in $\left(  \ref{order-of-mag}\right)  $%
.\vspace{0.3in}

Lastly, we return to $\left(  \ref{Byers-Yang in curved space-time}\right)  $
and develop an expression that correctly describes the quantization condition
for a superconducting ring in the presence of a charged \textquotedblleft mass
solenoid.\textquotedblright\ Rather than applying $\nabla^{-2}$ to both sides
of the gravito-Ampere law as DeWitt did in $\left(  \ref{del x h_0}\right)  $,
we can evaluate the gravito-magnetic flux directly as is done in $\left(
\ref{Flux in terms of rho}\right)  $ of Appendix B which gives $\Phi_{B_{G}%
}=\tfrac{1}{2}\mu_{G}\pi R^{4}\rho_{m}\omega$. Similarly, the magnetic flux
can be found using $\mu_{G}\rightarrow-\mu_{0}$ and $\rho_{m}\rightarrow
\rho_{c}$ which gives $\Phi_{B}=-\tfrac{1}{2}\mu_{0}\pi R^{4}\rho_{c}\omega$.
Also using $\left(  \ref{E_G}\right)  $, $\left(  \ref{h and B_G}\right)  $,
and $\left(  \ref{A and B}\right)  $ in $\left(
\ref{Byers-Yang in curved space-time}\right)  $ gives%
\begin{equation}
\dfrac{m_{e}}{en_{s}}\oint\limits_{C}\vec{J}\mathcal{\cdot}d\vec{l}+F=nh
\end{equation}
where%
\begin{equation}
F\equiv\pi R^{4}\omega\left(  4m_{e}\mu_{G}\rho_{m}-e\mu_{0}\rho_{c}\right)
-\frac{\pi eR^{6}\omega\rho_{m}\rho_{c}}{2\varepsilon_{G}\varepsilon_{0}c^{4}}%
\end{equation}
This result demonstrates that the persistent current is quantized in units of
$nh$ with an offset given by $F$ due to the presence of a charged mass
solenoid.\pagebreak

\noindent\underline{\textbf{Appendix A: The spatial inverse metric}}\bigskip

The \textquotedblleft spatial inverse metric\textquotedblright\ can be defined
as $\tilde{g}^{ik}$ where%
\begin{equation}
\tilde{g}^{ik}g_{jk}=\delta_{~j}^{i} \label{g^ik tilda}%
\end{equation}
To find an expression for $\tilde{g}^{ik}$, we can develop relations between
the metric and inverse metric components. Since $g^{\mu\nu}g_{\lambda\nu
}=\delta_{\lambda}^{\mu}$, then summing over $\nu$ gives%
\begin{equation}
g^{\mu0}g_{\lambda0}+g^{\mu k}g_{\lambda k}=\delta_{~~\lambda}^{\mu}
\label{g_mu,nu*g^mu,nu}%
\end{equation}
We can consider the various combinations of choosing space and time components
for $\mu$ and $\lambda$. Using $\left(  \ref{g_mu,nu*g^mu,nu}\right)  $ and
recognizing that $\delta_{~i}^{i}=\delta_{~0}^{0}=1$ and $\delta_{~0}^{i}=0$,
gives the following.%
\begin{align}
& \nonumber\\
\text{For }\left(  \mu,\lambda\right)   &  =\left(  i,j\right)  \hspace
{0.2in}\Longrightarrow\hspace{0.2in}g^{i0}g_{j0}+g^{ik}g_{jk}=\delta_{~j}%
^{i}\label{spatial kronecker}\\
& \nonumber\\
\text{For }\left(  \mu,\lambda\right)   &  =\left(  0,j\right)  \hspace
{0.2in}\Longrightarrow\hspace{0.2in}g^{00}g_{j0}+g^{0k}g_{jk}=\delta_{~j}%
^{0}\hspace{0.2in}\Longrightarrow\hspace{0.2in}g^{0k}g_{jk}=-g^{00}%
g_{j0}\label{g_jk}\\
& \nonumber\\
\text{For }\left(  \mu,\lambda\right)   &  =\left(  j,0\right)  \hspace
{0.2in}\Longrightarrow\hspace{0.2in}g^{j0}g_{00}+g^{jk}g_{0k}=\delta_{~0}%
^{j}\hspace{0.2in}\Longrightarrow\hspace{0.2in}g^{jk}g_{0k}=-g^{j0}%
g_{00}\label{g^jk*g_0k}\\
& \nonumber\\
\text{For }\left(  \mu,\lambda\right)   &  =\left(  0,0\right)  \hspace
{0.2in}\Longrightarrow\hspace{0.2in}g^{00}g_{00}+g^{0k}g_{0k}=\delta_{~0}%
^{0}\hspace{0.2in}\Longrightarrow\hspace{0.2in}g^{0k}g_{0k}=1-g^{00}%
g_{00}\qquad\label{g^00*g_00}\\
& \nonumber
\end{align}
Inserting $\left(  \ref{spatial kronecker}\right)  $ into $\left(
\ref{g^ik tilda}\right)  $ and dividing by $g_{jk}$ gives $\tilde{g}%
^{ik}=g^{ik}+g^{i0}g_{j0}/g_{jk}$. From $\left(  \ref{g_jk}\right)  $ we also
have $g_{jk}=-g^{00}g_{j0}/g^{0k}$ which leads to%
\begin{equation}
\tilde{g}^{jk}=g^{ik}-\dfrac{g^{0i}g^{0k}}{g^{00}} \label{g^tilda}%
\end{equation}
This expression is valid to \textit{all} orders in the metric. If we use
$g_{\mu\nu}=\eta_{\mu\nu}+h_{\mu\nu}$ and $g_{\mu\sigma}g^{\sigma\nu}%
=\delta_{\mu}^{\hspace{0.05in}\nu}$, then the inverse metric (to first order
in the metric perturbation) is $g^{\mu\nu}=\eta^{\mu\nu}-h^{\mu\nu}$. Also
using $\left(  1+h^{00}\right)  ^{-1}\approx1-h^{00}$ leads to%
\begin{equation}
\tilde{g}^{jk}\approx g^{ik}+h^{0i}h^{0k}\left(  1-h^{00}\right)
\label{g^tilda_jk to first order}%
\end{equation}
Therefore, $\tilde{g}^{jk}\approx g^{ik}$ is only true to first order in the
metric perturbation.\bigskip

There are two other quantities that appear in the Hamiltonian in $\left(
\ref{H_general}\right)  $ that can be evaluated here. The first quantity is
$\tilde{g}^{ik}g_{0k}$. Writing $\left(  \ref{g_jk}\right)  $ with
$\mu=0,~\lambda=k$ and using $j$ for the repeated index gives $g_{jk}%
=-g^{00}g_{k0}/g^{0j}$. Inserting this into $\left(  \ref{g^ik tilda}\right)
$ leads to%
\begin{equation}
\tilde{g}^{ik}g_{0k}=-\dfrac{g^{0i}}{g^{00}} \label{g^ik*g_0k}%
\end{equation}
The second quantity is $\left(  \tilde{g}^{jk}g_{0j}g_{0k}-g_{00}\right)  $
which can be evaluated by using $\left(  \ref{g^tilda}\right)  $. This gives%
\begin{align}
\tilde{g}^{jk}g_{0j}g_{0k}-g_{00}  &  =g^{jk}g_{0j}g_{0k}-\dfrac{g^{0j}g^{0k}%
}{g^{00}}g_{0j}g_{0k}-g_{00}\\
& \nonumber
\end{align}
Using $\left(  \ref{g^jk*g_0k}\right)  $ and $\left(  \ref{g^00*g_00}\right)
$ leads to%
\begin{equation}
\tilde{g}^{jk}g_{0j}g_{0k}-g_{00}=-\dfrac{1}{g^{00}}
\label{g^tilda^jk*g_oj^g_ok}%
\end{equation}
\vspace{0.3in}\vspace{0.3in}

\noindent\underline{\textbf{Appendix B:\ Gravito-electromagnetic fields of a
non-relativistic \textquotedblleft mass solenoid\textquotedblright}}\bigskip

Here we consider a rotating cylindrical \textquotedblleft mass
solenoid\textquotedblright\ of length $\ell$ and radius $R$ (where $\ell>>R$)
with the axis along the $z$-axis from $z=-\ell/2$ to $z=\ell/2$. We assume
that the cylinder rotates at a constant non-relativistic angular velocity and
hence has a non-time varying mass current.%

\begin{figure}[h]%
\centering
\includegraphics[
trim=-0.277075in 0.000000in -0.234448in -0.127356in,
natheight=7.236100in,
natwidth=6.875300in,
height=3.4843in,
width=3.4973in
]%
{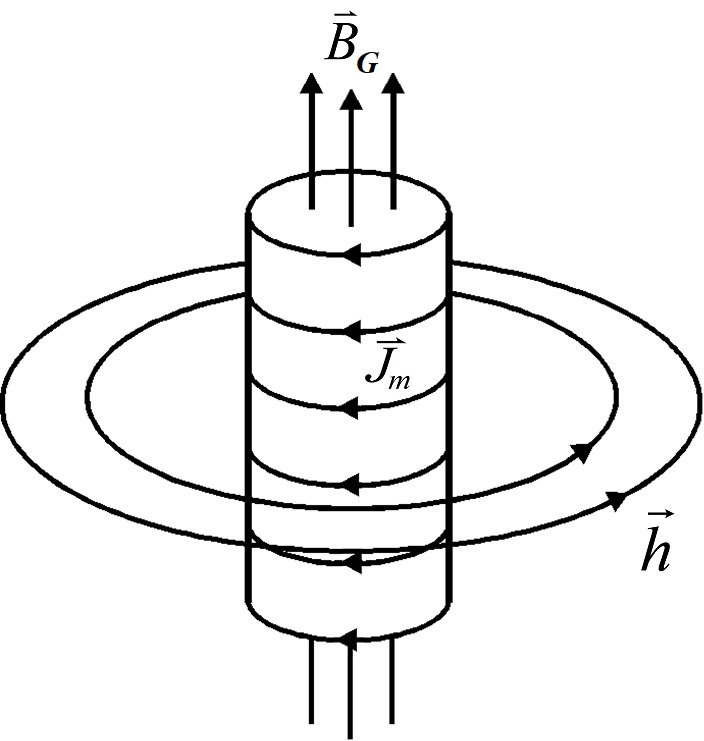}%
\caption{A \textquotedblleft mass solenoid\textquotedblright\ with a mass
current, $\vec{J}_{m}$, creating a gravito-vector potential, $\vec{h}$, and
corresponding gravito-magnetic field, $\vec{B}_{G}$. Note that in the diagram,
$\vec{h}$ points in the \textit{opposite} direction of $\vec{J}_{m}$ as a
result of the negative sign in $\nabla\times\left(  \nabla\times\vec
{h}\right)  =-\mu_{G}\vec{J}_{m}$.}%
\end{figure}

We can use the gravito-Gauss law from $\left(  \ref{Gauss and Ampere}\right)
$ to obtain the gravito-electric field, $\vec{E}_{G}$. Taking the volume
integral of both sides and applying the Divergence theorem gives%
\begin{equation}%
{\displaystyle\oint\limits_{\substack{\text{\textit{Surface}}%
\\\text{\textit{of V}}}}}
\vec{E}_{G}\cdot d\vec{A}=-\dfrac{1}{\varepsilon_{G}}\int\limits_{V}\rho_{m}dV
\end{equation}
Assume a uniform mass distribution and use a cylindrical Gaussian surface with
radius $r$ surrounding the mass solenoid concentrically. This gives%
\begin{equation}
\vec{E}_{G}=-\dfrac{R^{2}\rho_{m}}{2\varepsilon_{G}r}\hat{r} \label{E_G}%
\end{equation}
We can also use a line integral of $\vec{E}_{G}$ to find the change in the
gravito-scalar potential, $\varphi_{G}$, when going from $r^{\prime}=r_{0}$ to
$r^{\prime}=r$, where $r_{0}$ is an arbitrary distance away from the mass
solenoid such that $\varphi_{G}\left(  r_{0}\right)  =0$.%
\begin{equation}
\Delta\varphi_{G}\left(  r\right)  =-\int_{r_{0}}^{r}\vec{E}_{G}\cdot d\vec{r}%
\end{equation}
Evaluating the integral gives%
\begin{equation}
\varphi_{G}\left(  r\right)  =\dfrac{R^{2}\rho_{0}}{2\varepsilon_{G}}%
\ln\left(  \dfrac{r}{r_{0}}\right)  \label{Scalar potential for solenoid}%
\end{equation}

Next we can calculate the gravito-vector potential, $\vec{h}$, outside the
mass solenoid. Since the z-axis is the axis of symmetry as well as the axis of
rotation, then $\vec{h}=h_{\phi}\left(  \vec{r}\right)  \hat{\phi}$. To find
an expression for $\vec{h}$, we take a line integral of $\vec{h}$ along a
closed path around the solenoid, use $\vec{B}_{G}=\nabla\times\vec{h}$ and
apply Stokes' theorem.%
\begin{equation}%
{\displaystyle\oint\limits_{\substack{\text{\textit{Around}}%
\\\text{\textit{solenoid}}}}}
\vec{h}\cdot d\vec{r}=\int\limits_{\substack{\text{\textit{Cross section}%
}\\\text{\textit{of solenoid}}}}\left(  \nabla\times\vec{h}\right)  \cdot
d\vec{S}=\int\limits_{\substack{\text{\textit{Cross section}}%
\\\text{\textit{of solenoid}}}}\vec{B}_{G}\cdot d\vec{S}=\Phi_{\vec{B}_{G}}
\label{gravitational flux}%
\end{equation}
where $\Phi_{B_{G}}$ is the total gravito-magnetic flux of $\vec{B}_{G}$
through a cross-section of the solenoid. If we use a circular path along the
$\hat{\phi}$ direction (with $z$ in the upward direction), then we also have%
\begin{equation}%
{\displaystyle\oint\limits_{\substack{\text{\textit{Around}}%
\\\text{\textit{solenoid}}}}}
\vec{h}\cdot d\vec{r}=h_{\phi}2\pi r \label{Line integral of h}%
\end{equation}
So equating $\left(  \ref{gravitational flux}\right)  $ and $\left(
\ref{Line integral of h}\right)  $\ gives%
\begin{equation}
\vec{h}=\dfrac{\Phi_{B_{G}}}{2\pi r}\hat{\phi}
\label{h_vector in terms of B_g flux}%
\end{equation}

We can also develop an expression for $\Phi_{B_{G}}$ (and hence for $\vec{h}$)
by taking a surface integral of both sides of the gravito-Ampere law from
$\left(  \ref{Gauss and Ampere}\right)  $ and applying Stokes' theorem to
change the surface integral into a line integral which gives%
\begin{equation}%
{\displaystyle\oint}
\vec{B}_{G}\cdot d\vec{l}=-\mu_{G}I_{m}%
\end{equation}
where $I_{m}=\int\vec{J}_{m}\cdot d\vec{S}$ is the mass-current. We can use a
line integral along a rectangular loop with one edge \textit{inside} the
solenoid parallel to the axis (where $\vec{B}_{G}\not =0$) and the opposite
edge \textit{outside} the solenoid (where $\vec{B}_{G}=0$). If the length of
the edge is $L$, then we obtain%
\begin{equation}
B_{G}L=-\mu_{G}I_{m}%
\end{equation}
The total current in a solenoid is $I_{m}=Ni_{m}$ where $i_{m}$ is the current
in each loop. If the solenoid is a continuous mass shell, then it is
effectively a \textquotedblleft perfect\textquotedblright\ solenoid where the
current is distributed continuously over the surface. Then we can use
$J_{m}=\sigma_{m}\omega$ where $\sigma_{m}$ is the effective surface mass
density of the cylinder spinning with angular velocity $\omega$. So the total
current would be $I_{m}=J_{m}A_{\bot}$ where $A_{\bot}=RL$ is the area normal
to the current, and $R$ is the radius of the solenoid. Then we have
$B_{G}L=-\mu_{G}\left(  \sigma_{m}\omega\right)  \left(  RL\right)  $ which
gives%
\begin{equation}
B_{G}=-\mu_{G}R\sigma_{m}\omega\label{B_g}%
\end{equation}
We can now determine the magnitude of the gravito-magnetic flux $\Phi_{B_{G}}$
through a cross-sectional area of the solenoid, $A_{cs}=\pi R^{2}$. When we
determined the gravito-magnetic field, we already treated the cylinder as a
\textquotedblleft perfect\textquotedblright\ solenoid which means it is
effectively one \textquotedblleft loop\textquotedblright\ so $N=1$. Then we
have%
\begin{equation}
\Phi_{B_{G}}=NB_{G}A_{cs}=\left(  \mu_{G}R\sigma_{m}\omega\right)  \left(  \pi
R^{2}\right)  =\mu_{G}\pi R^{3}\sigma_{m}\omega\label{gravito-magnetic flux}%
\end{equation}
We can also express this in terms of a \textit{volume} mass density since%
\begin{equation}
\sigma_{m}A=\rho_{m}V\qquad\implies\qquad\sigma_{m}\left(  2\pi RL\right)
=\rho_{m}\left(  \pi R^{2}L\right)  \qquad\implies\qquad\sigma_{m}=R\rho_{m}/2
\end{equation}
Then the gravito-magnetic flux in $\left(  \ref{gravito-magnetic flux}\right)
$ becomes%
\begin{equation}
\Phi_{B_{G}}=\tfrac{1}{2}\mu_{G}\pi R^{4}\rho_{m}\omega
\label{Flux in terms of rho}%
\end{equation}
We can substitute this back into $\left(  \ref{h_vector in terms of B_g flux}%
\right)  $ and $\left(  \ref{B_g}\right)  $ to express the gravito-vector
potential and the gravito-magnetic field in terms of the physical parameters
of the mass solenoid. This gives, respectively,%
\begin{equation}
\vec{h}=\dfrac{\mu_{G}R^{4}\rho_{m}\omega}{4r}\hat{\phi}\qquad\text{and}%
\qquad\vec{B}_{G}=-\tfrac{1}{2}\mu_{G}R^{2}\rho_{m}\omega\hat{z}
\label{h and B_G}%
\end{equation}
If the mass solenoid is also charged, then there will also be a magnetic
vector potential and magnetic field. These can be found by simply comparing
$\nabla\times\vec{B}_{G}=-\mu_{G}\vec{J}_{m}$ and $\nabla\times\vec{B}=\mu
_{0}\vec{J}_{c}$ which implies $\mu_{G}\rightarrow-\mu_{0}$ and $\rho
_{m}\rightarrow\rho_{c}$. Then $\left(  \ref{h and B_G}\right)  $ can be used
to immediately obtain%
\begin{equation}
\vec{A}=-\dfrac{\mu_{0}R^{4}\rho_{c}\omega}{4r}\hat{\phi}\qquad\text{and}%
\qquad\vec{B}=\tfrac{1}{2}\mu_{0}R^{2}\rho_{c}\omega\hat{z} \label{A and B}%
\end{equation}
Lastly, the electric field can also be obtained by using $\varepsilon
_{G}\rightarrow-\varepsilon_{0}$ and $\rho_{m}\rightarrow\rho_{c}$ in $\left(
\ref{E_G}\right)  $ to obtain%
\begin{equation}
\vec{E}=\dfrac{R^{2}\rho_{c}}{2\varepsilon_{0}r}\hat{r} \label{E}%
\end{equation}

\end{document}